\documentclass[]{interact}

\usepackage{epstopdf}% To incorporate .eps illustrations using PDFLaTeX, etc.
\usepackage[caption=false]{subfig}% Support for small, `sub' figures and tables
\usepackage{amsmath, amsfonts, amssymb} % math equations, symbols
\usepackage[english]{babel}
\usepackage{color}      % color content
\usepackage{graphicx}   % import figures
\usepackage{url}        % hyperlinks
\usepackage{bm}         % bold type for equations
\usepackage{multirow}
\usepackage{booktabs}
\usepackage{epstopdf}
\usepackage{epsfig}
\usepackage{algorithm}
\usepackage{algorithmicx}
\usepackage{titlesec}
\usepackage{booktabs}
\usepackage{algpseudocode}
\usepackage{amsmath}
\usepackage{xcolor}
\usepackage{tikz}
\usepackage{amssymb}

\usepackage{multicol}
\usepackage{enumerate}
\usepackage{hyperref}
\usepackage{algorithm,algpseudocode,float}
\usepackage{lipsum}
\usepackage{amsmath}
\usepackage[utf8x]{inputenc}
\usepackage[titletoc]{appendix}
\usepackage{mathrsfs}
\usepackage{amsmath,amsthm,amsfonts,amssymb,amscd}



\theoremstyle{plain}% Theorem-like structures provided by amsthm.sty

\theoremstyle{definition}

\theoremstyle{remark}

\begin{document}

\title{Drug dissemination strategy with an SEIR-based SUC model}

\author{
\name{Boyue Fang\textsuperscript{a}\thanks{Email: byfang97@gmail.com} and Yutong Feng\textsuperscript{b}}
\affil{\textsuperscript{a}School of Mathematical Sciences, Fudan University ; \textsuperscript{b}Electronics Science and Technology, Fudan University}
}

\maketitle
  \renewcommand{\baselinestretch}{1.5} \selectfont
  \begin{abstract}
According to the features of drug addiction, this paper constructs an SEIR-based SUC model to describe and predict the spread of drug addiction. Predictions are that the number of drug addictions will continue to fluctuate with reduced amplitude and eventually stabilize. To seek the fountainhead of heroin, we identified the most likely
origins of drugs in Philadelphia, PA, Cuyahoga and Hamilton, OH, Jefferson, KY, Kanawha, WV, and Bedford, VA. Based on the facts, advised concentration includes the spread of Oxycodone, Hydrocodone, Heroin, and Buprenorphine. In other words, drug transmission in the two states of Ohio and Pennsylvania require awareness. According to the propagation curve predicted by our model, the transfer of KY
state is still in its early stage, while that of VA, WV is in the middle point, and OH, PA in its latter ones. As a result of this, the number of drug addictions in KY, OH, and VA is projected to increase in three years. For methodology, with the Principal component analysis technique, 22 variables in socio-economic data related to the continuous use of Opioid drugs was filtered, where the 'Relationship' Part deserves a highlight.
Based on them, by using the K-means algorithm, 464 counties were categorized into three baskets. To combat the opioid crisis, a specific action will discuss in the sensitivity analysis
section. After modeling and analytics, innovation is required to control addicts and advocate anti-drug news campaigns. This part also verified the effectiveness of model when $d_1<0.2; r_1,r_2,r_3<0.3; 15<\beta_1,\beta_2,\beta_3<25$. In other words, if such boundary exceeded, the number of drug addictions may rocket and peak in a short period.
 \end{abstract}
 \section{Introduction}
 \subsection{Background}
 Opioid, the substance that produces morphine-like effects on receptors, spreads widely throughout America. Although medically used for pain relief in some prescriptions, such as anesthesia, opioids impose side effects associated with vomit and constipation. In other words, those who take opioids are more vulnerable.

On the other hand, opioids do not entail specific organ toxicity. Unlike other drugs such as aspirin and paracetamol, it not even bounds up with kidney toxicity. However, the immune system of opioid takers has collapsed.

If people spend huge money to purchase opioids, direct public spending will increase significantly, because they may lose work because of the addition. In other words, the rate of crime, unemployment, and homelessness will increase, which means burdening the national fiscal budget to combat corruption and raise social welfare relief scale.
Furthermore, if the opioid crisis affects all classes of American society, country stability may come under challenge. People with a higher education background, honorable occupation, and higher social status are associated with the well-being of the entire economy, especially those who work for complex businesses that require precise labor skills.

Hence, strategy to face the opioid abuse based on the records and tendency appears urgent. This essay highlights the issues mentioned above and intends to advise possible solutions with mathematical models and visual graphics.
\subsection{Problem restatement}
Opioid abuse spread greatly over the years from 2010 to 2017 in counties of the five states. Under the situation, the question concentrates on the propagation mechanism. For a system of dissemination, the first thing is the input and the output, which will be further discussed in the following pages. In this issue, counties can be considered as nodes, and what is required to seek is the methodology to depict the spreading flow. Each flow has its trend (i.e., either increasing or decreasing), relationship and other correspondence, and possibly a source. Some may not have an exact origin, but the echo and resonance in and between also convey information that may be of use and implement the story. Therefore, the essential section of this paper falls on the insight. In other words, what the model shapes matters.
\section{Model preparedness}
\subsection{ Literature Review}
 Since White,E. et al.\cite{1}, ordinary differential equations (ODE) has been introduced for the heroin addiction model. They classified drugsters aging from 15 to 64 according to their record of treatment acceptance, with standard incidence rate(SIR) as the effective contact rate, and identified that when $R_0<1$, the no-drug equilibrium stabilizes gradually and eradicates. However, when $R_0>1$, there exists a balance point that is, however, toxic. Thus, Muroya,Y. et al. \cite{2} replaced SIR coefficient $\frac{\beta U_1U_2}{N}$ with linear term $\sigma U_2$ and set $R_0$ as the threshold to simulate the systematic global dynamic behavior. It is worth mention that addicts normally get addicted not because of one single drug case, but multiple, and based on the fact, modified the occurrence frequency to $\beta S^p$ . Contributors also considered a variety of media reporting factors. For example, Xiao,D. et al.\cite{3} introduced function $g_S(I)=\frac{kSI}{1+\alpha I^2}$ in SIR model to represent psychological suppression effect, isolating $R_0$ from $\alpha$, while I (parameter) still reciprocal to $\alpha$. Sun,C. et al.\cite{4}, Cui, J. et al. \cite{5}, Liu, Y. et al.\cite{6}, Tchuenche, j. M. et al.\cite{7} also discussed this problem within similar models.

Overall, Xiao, Y. et al.\cite{8} used a piecewise smooth incidence rate curve $\beta e^{\{-M(I,dI/dt)\}}$ as the propagation coefficient, in which $M(I,\frac{dI}{dt})=\max\{0,p_1 I(t)+p_2 \frac{dI(t)}{dt} \}$. This concludes that media or psychological factors delay the transmission peak and restrict the crisis to break out in a smaller range, but they cannot influence the propagation threshold. Liu, R. et al.\cite{9} categorized the crowd into susceptible(E), infective (I) and medical consumer(H) as an EIH model, using $\beta_0=\beta exp{\{-\alpha_1E-\alpha_2I-\alpha_3H\}}$ as the propagation coefficient. They noted that media reports might cause continuous periodic oscillations. Sahu, G. P. et al.\cite{10} also developed a non-linear SEQIHRS model, assumed the incidence rate as $\beta exp{\{\frac{m_1 I+m_2 H}{H}\}}$, in which I, H stand for infective and quarantined and confirmed that media report could alleviate the infection and transmission of drug abuse by reducing the number of the infective in the equilibrium state.
\subsection{Basic Assumptions}
   Note that the model is based on the similitude between the process of drug dissemination and the epidemic spread network, that

Both Epidemic and Drug Dissemination
\begin{itemize}
\item     Starts from one or more source
\item     Occurrence of drug report may equivalent to infection of epidemic
\item     Disseminates with similar behavior
\end{itemize}

For each area
\begin{itemize}
\item     Epidemic may change the area from infection to no-infection
\item     Epidemic may change the area from no-infection to infection
\item     Drug dissemination may change the area from NO-Drug-area to Drug-reported-area
\item     Drug dissemination may change the area from Drug-reported-area to NO-Drug-area
\end{itemize}

   \subsection{Model Landscape}
   The model analyzes the patterns and features of the spread of opioid abuse, and give the US government suggestions to combat the opioid crisis.

For opioid abuse, time can be considered as a variable in the propagation function. Therefore, based on the location and possible flow of states and counties, the purpose of the model is to ascertain trends and possible correlations. For example, by visualizing the source of the drug, or by filtering, the characteristics of these drugs can be obtained. The model that combined visualization and mining, including sensitivity analysis, tests the robustness and scope of the slight changes in conditions and known values. At the same time, through principal component analysis, find possible sources of abuse of opioids.

   \subsection{Notation}
   
The variables are as listed:

$S_i$         Indicates the number of addicts in the i-type county;

$U_{1i}$  Indicates the number of unexplained addicts in the i-type county;

$U_{2i}$  Indicates the number of addicts found in the i-type county;

$C_i$   Indicates the number of new addicts in the i-type county (active addiction);

$\alpha_{ki} $ Represents the addicts of the i-type county the addicts (unexplained), the influence factors of the addicts (found), which affect the number of addicts per year;

$\alpha_i $   Conversion rate of i-type county residents from not addicted to addicts;

$\beta_i $    The coefficient of influence of the influence of addicts on the i-type county addicts;

$d $ Natural mortality;

$d_1$   Drug user mortality;

$h_i $    Proportion of addicts in the i-th county from unexplained to found;

$r$  Proportion of successful detoxification of the county dwellers;

$s$     Initial value, $s=(s_1,s_2,s_3,u_{11},u_{12},u_{13},u_{21},u_{22},u_{23} )$.

 \section{The models}
 \subsection{Part 1: SUC Model}
 \subsubsection{Model assumption}
 1. Suppose that there are 4 types of people within the spreading scope of opioid.

i. Susceptible. (denote as S), who is not addicted yet, but is likely to be exposed to opioids, and become addicted.

ii. Addict, but undiscovered(denote as $U_1$), who is addicted but not detected by the outside world.

iii. Addict, and discovered(denote as $U_2$), who is addicted and detected by the outside world.

iv. Difficult to be addicted, due to reasons such as the restriction of age.

2. Suppose the relative proportion of susceptible(A), addict(B), and people difficult to be addicted(C) remains the same in all parts of the population. The net conversion rate from C to A because of his factor is a constant(denote as c). That because the impact of the other three kinds of people is f(S,$U_1$,$U_2$).

3. The part of the population of susceptible transformed into addicts is denoted as $\alpha^*$ stand they will not be detected at first.

4. The death rate of people who aren't addicted and addicts are both constants and are denoted as $d$ and $d_1$ respectively.

5. The probability of $U_1$ being detected is a constant $h$, and once detected, it becomes $U_2$ without delay.

6. Undiscovered addicts won't receive drug treatment on their own. The proportion of addicts discovered that receive drug treatment is a constant, denoted as $r$.

7. New birth and the migration of residents are not taken into consideration.

   \subsubsection{SUC Model construction}
   The following differential equations represent the state function of the model.
   \begin{eqnarray}
   \left\{
   \begin{array}{lll}
   \frac{dS}{dt} = b-f(U_1,U_2)S -dS+\delta U_2 \\
   ~\\
   \frac{dU_1}{dt} = f(U_1,U_2)S -vU_1+\sigma U_2 -d_1U_1\\
   ~\\
   \frac{dU_2}{dt} = vU_1-\sigma U_2 -d_2U_2 -\delta U_2. \\
   \end{array}{}
   \right.
   \end{eqnarray}

   Why we build the model based on the epidemic model? Firstly, there is a similarity between drugs and the epidemic that once people are infected with them, it is difficult to cure. And it has been discussed in the classic SIR model and its derivative models that people with infectious diseases may be healed and infected with it again, which is very similar to a drug relapse. Similarly, the prevention and control work is essential but easy to be neglected.

Moreover, judging from the cause of drug addiction, most people become addicted because of other people's cheating, which is similar to the theory of the spread of the epidemic. But we must notice that addiction due to the personal factor needs to be taken into consideration in the case of drug addiction. It is one of their main differences, which we will discuss later.

To pay attention to the prevention and restriction of the spread of drugs, we focus on the come into being susceptible. So we set the influence function $f$ as a constant function, and the factor concerning the transformation of ordinary people into susceptible as an exponential function. Here we take $b= \beta exp\{-S(\alpha_1 S+\alpha_2 U_1+\alpha_3 U_3 ) \}$, where $\alpha_i$ is constant.

From these ordinary differential equations, initialize the original values and plot the curve in a three-dimensional grid, presents the wall of the fluctuation.

      \begin{figure}[H]
      \centering
      \includegraphics[width=.8\textwidth]{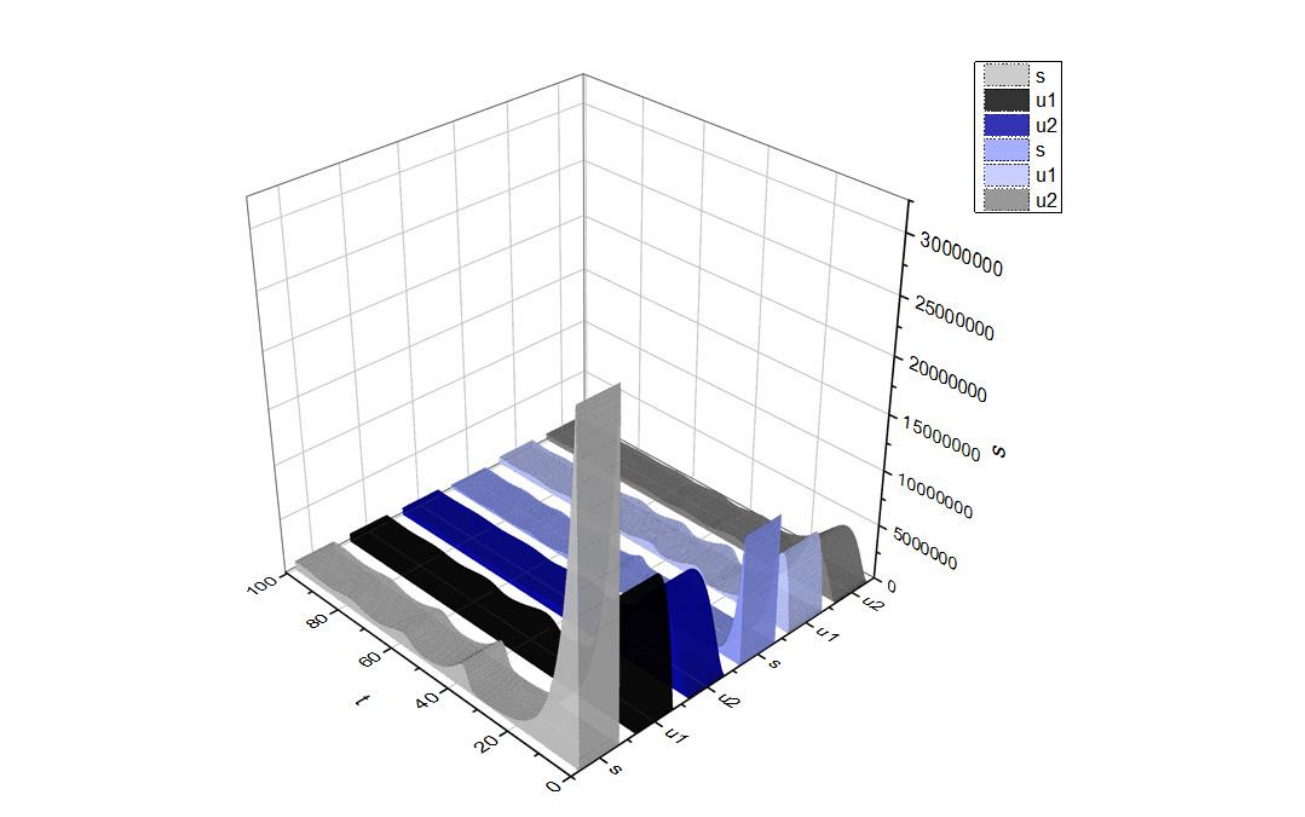}
      \caption{Curve numerical plot}
      \end{figure}

     To each non-analytic solution, the following diagram gives a plot of the numerical solutions and the scale.

\begin{figure}[H]
      \centering
      \includegraphics[width=.8\textwidth]{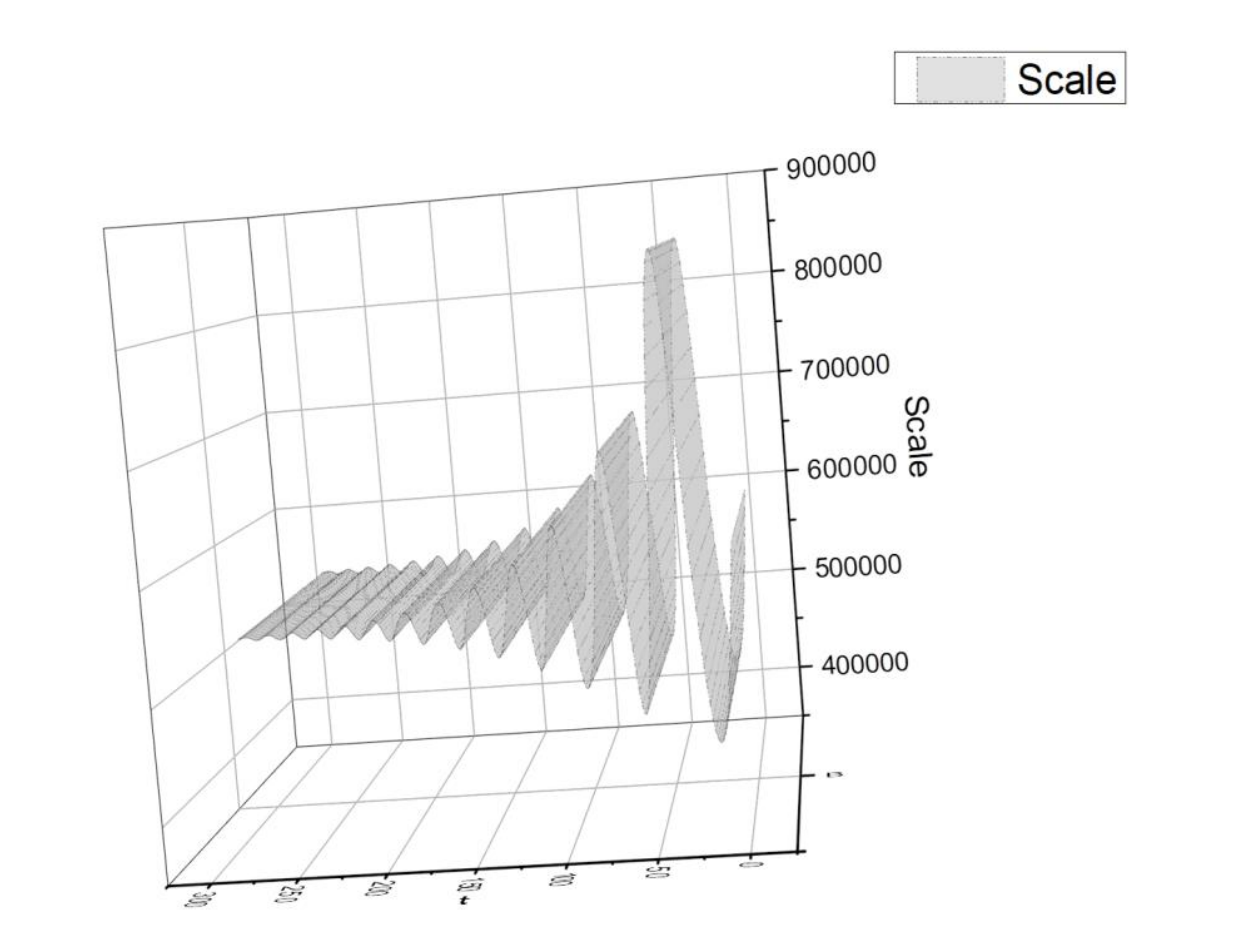}
      \caption{Solution numerical plot}
      \end{figure}

     Concerning the threshold value of drug identification, we considered the minimum number of generations. Liu, R. et al.\cite{9} approximated the minimum number of generation by $\frac{\beta}{5(d+h)} \approx 20$. It means although we optimistically adjust $\alpha_2$, $\alpha_3$ to a shallow level, as long as an addict influences more than 20 people, transforming them into susceptible persons. The number of addicts will continuously grow. Fortunately, we are not at this stage yet.

After having developed the model, we can find that based on the assumptions above, we come to a startling conclusion that the fluctuation of the number of addicts follows the trend of vibration. In judging whether this result corresponds with the actual situation of the five states, we consider correlation analysis.

After preprocessing data and getting numerical values from the model (namely SUC, from the parameters used), curves yield a predictive perspective of the propagation of the wave.

Under the operation of the model, number of people who are drug addicted and discovered fluctuates, which accords with the actual situation. This get further confirmed from the testing results of the given data as follows:

\begin{table}[H]
\begin{tabular}{@{}llllll@{}}
\toprule
Position               & KY  & OH  & PA  & VA  & WV  \\ \midrule
2010                   & 51  & 138 & 172 & 106 & 100 \\
2011                   & 52  & 139 & 173 & 112 & 101 \\
2012                   & 53  & 140 & 174 & 118 & 102 \\
2013                   & 54  & 141 & 175 & 124 & 103 \\
2014                   & 55  & 142 & 176 & 130 & 104 \\
2015                   & 56  & 143 & 177 & 136 & 105 \\
2016                   & 57  & 144 & 178  & 142 & 106 \\
2017                   & 58  & 145 & 179 & 148 & 107 \\
Correlation coefficient & 0.90 & 0.99 & 0.90 & 0.89 & 0.97 \\ \bottomrule
\end{tabular}
     \caption{Correlation coefficient by state 2010 - 2017}
\end{table}

We can find out that these two sets of data have a strong correlation, and the circumstances of the spread of opioids in the five states are different from one another.

As far as the current situation is concerned, according to the propagation curve predicted by our model, the transmission of KY state is in the early stage, the transfer of VA, WV state is in the medium term, the communication of OH, PA state is in the late stage, and we predict the number of drug addictions in KY, OH, and VA will increase in three years, with a particular focus on OH.

Meanwhile, as the model states, viewing from a visualized angle, we again notice that the NO-Drug counties vibrate less and less intense and eventually reach an equilibrium during the process of emerging then perish. If the same pattern and characteristics of the spread of the opioid abuse continue, the increase of addicts will worsen social stability and increase the medical burden.

    \begin{figure}[H]
      \centering
      \includegraphics[width=.8\textwidth]{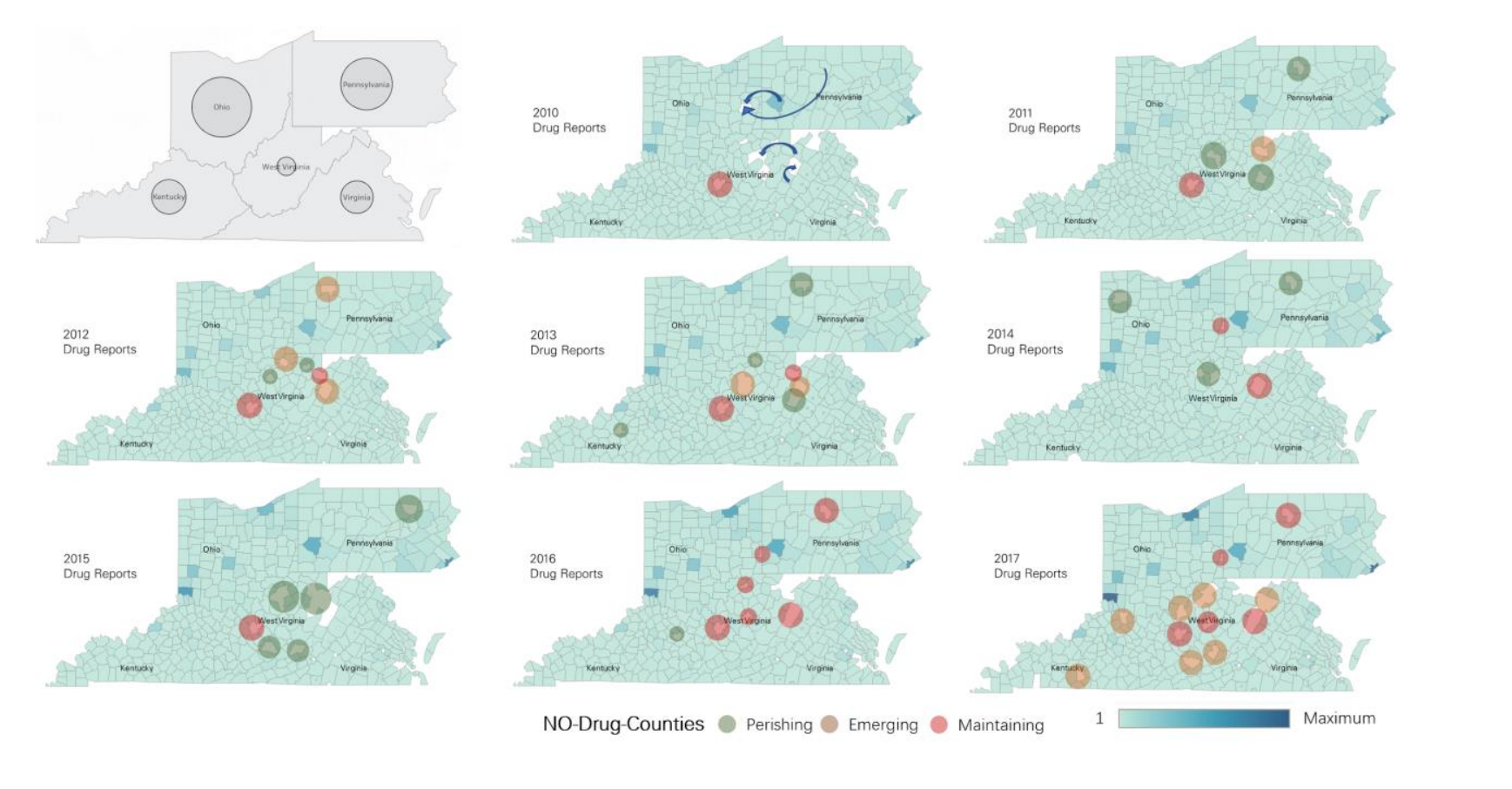}
      \caption{Change of No-drug counties}
    \end{figure}

     Moreover, findings are that under the assumptions of our model that many addicts remain to be undiscovered, the number of outnumbers that of the people who have already been discovered. In other words, It will be an outbreak at any time. Possible concerns of the US government also involve that if there appears an addictive drug with a broader influence, which surpasses the threshold value of the model, more and more people, even the majority will become opioid abusers.

   \subsubsection{Model explanation: track the source}

Based on our SUC Model, we transformed maps of every state into the morphological graph and obtained diagrams containing distances between each state and the number of drug reports. The following take heroin as an example. We deeply analyzed its propagation feature and infection source.

    \begin{figure}[H]
      \centering
      \includegraphics[width=.8\textwidth]{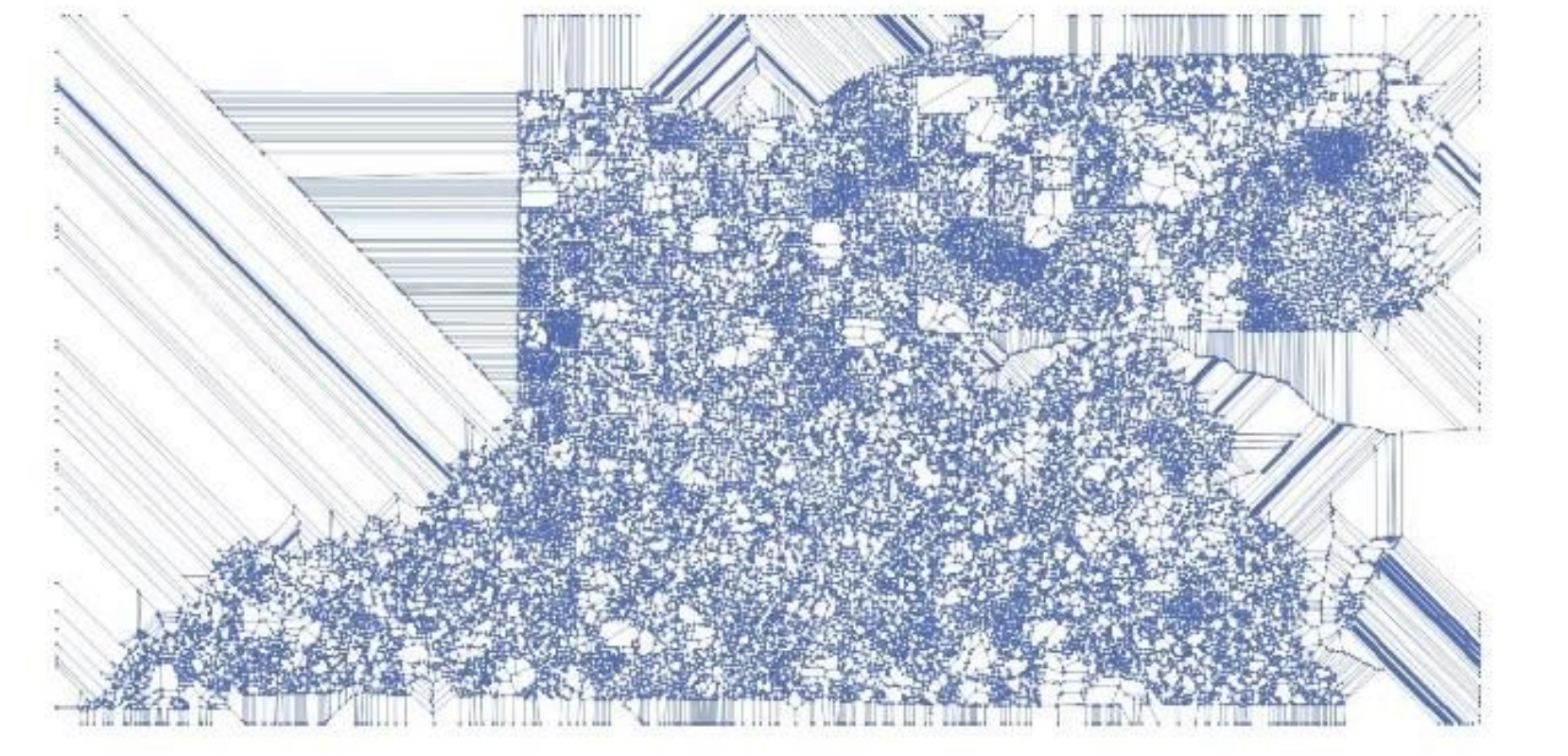}
      \caption{Morphological graph of counties}
    \end{figure}

The simple virus propagation model, a similar propagation simulation, was searched by Shad, D. et al. \cite{11} for the source of the virus as an issue of MLE(Maximum Likelihood Estimation). Denote the diagram above as G(V, E), where V is a finite point set, and E an edge set. Assume that the virus has already spread to and infected N nodes, which are shown in the figure. Denote the source of the infection as v*, we have the Maximum Likelihood Estimator: Equation Chapter (Next) Section 1Equation Section (Next)

   \begin{equation}
   v =\arg\max_{v \in G_N}\{\mathbb{P}(G_N | V^*=v)\}
   \end{equation}

   For heroin, as illustrated in the following charts, consistently occupied the top 3 drug reports county total from 2010 - 2017, and is popular among Ohio. Gradually climbing up and exceeds a total drug report of 200,000.

    \begin{figure}[H]
      \centering
      \includegraphics[width=.8\textwidth]{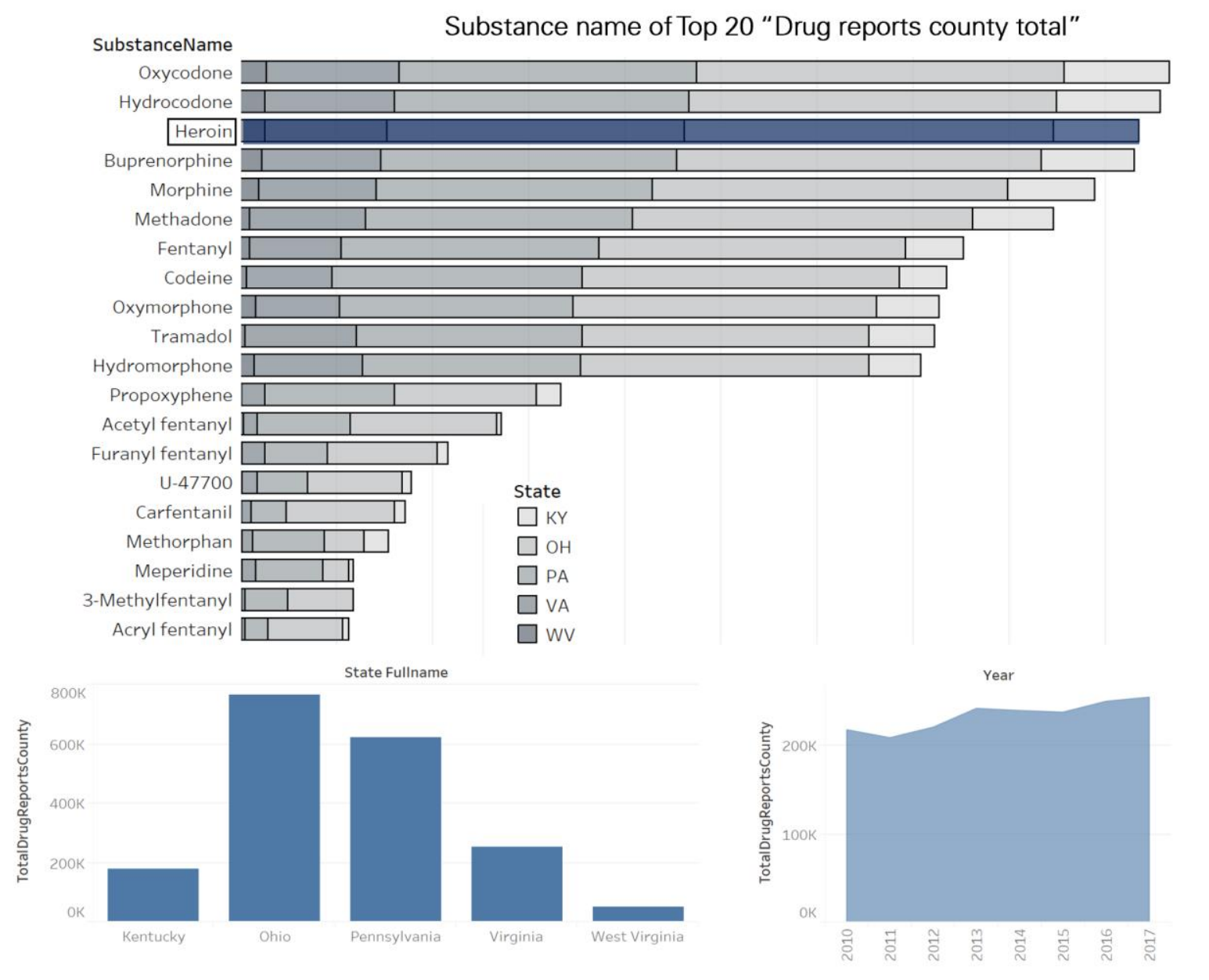}
      \caption{Heroin: substantial statistical landscape}
    \end{figure}

Findings according to the classifier and structural analysis are that there is a trend in the spread of heroin, showing a wider range of influence and a more significant number of heroin addicts. Moreover, by analyzing heroin separately and comparing its data to other opioids, we can find out that its impact surpasses the average level of opioids. Assuming the latent population of heroin addicts has the same features with that of other opioids, we can conclude that among the total population of addicts, a more substantial proportion of heroin addicts will lead to a more significant number of the total population of addicts, when the circumstance of the drug users reaches toxic equilibrium.

   \subsection{Part 2: SUC(I) Model}
   \subsubsection{Principal Component Analysis (PCA) Data processing }

   This section intends to take socio-economic factors into account. Before correlation filter, dimensionality reduction techniques, such as missing value ratio, Low Variance Filter, and factor analysis reduced the dimension to around 100.

Meanwhile, to consider the target as a whole, take the average of the 2010 - 2017 data as a new data frame.

The criteria can be seen below:
\begin{itemize}
\item     Remain dimension with name "percent" only;
\item     Remove meaningless dimensions;
\item     Adjust data frame shape: line of "Bedford City, Virginia" was eliminated from the list since 2014. Because according to the household size that is lower than the average of the list, we assume that removing the line will not affect the entire data greatly;
\item     Remove columns in the middle with [x] mark higher than 80\%;
\item     Remove dimensions with any top 80\% small values less than 5\%.
\end{itemize}

And the proliferation rules are:
\begin{itemize}
\item            Identify big city: set the average of the household as benchmark and mark as big city;
\item            Identify noise: For big cities, if the ratio of big city total is smaller than 5\% of the dimension total, then it is considered as anomalies;
\item            Remove noise: anomalies here are considered as noise, therefore remove such dimension.
\end{itemize}

Then consider the high correlation filter. For the remaining 90 parameters, plot the correlation matrix and apply the High correlation filter, we obtain 26 dimensions eventually.

Note that the following dimensions are highly correlated with many other factors:
\begin{itemize}
\item     Households with one or more people under 18 years;
\item     RELATIONSHIP - Population in households.
\end{itemize}
This indicates that the relationship is adding weight when predicting the trend with the built SUC model.

       \begin{figure}[H]
      \centering
      \includegraphics[width=.8\textwidth]{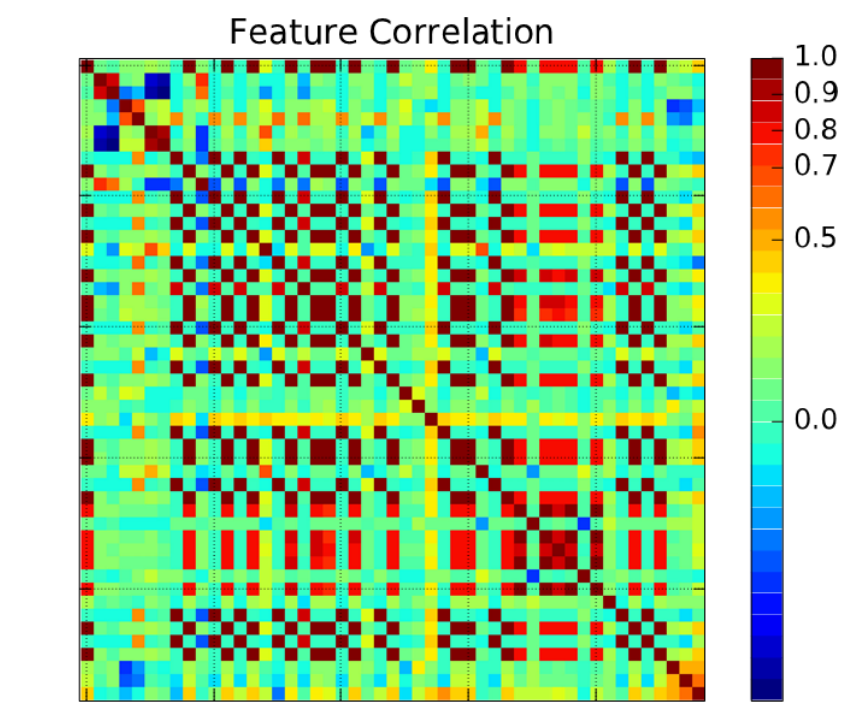}
      \caption{Correlation matrix}
      \end{figure}

     Based on 26 dimensions extracted by the PCA technique, apply the K-means algorithm, we may divide the counties into three types, based on the demographic data provided.

We consider that people living in these three kinds of counties have different characteristics, so we refine the assumptions raised by the SUC model. We think that the overall situation is like what the SUC model describes, and people living in these three types of counties have different possibilities of drug use.

     \subsubsection{SUC(I) Model construction}

By analyzing the population of economic data, we find out that with a different state of development, the probabilities of residents abusing opioids are different in these counties. So, we refined our model to include the socio-economic factors. After analyzing the annual data of each county, we divided all the counties into three kinds, and the susceptible and addicts into three types according to the country they belong to.

So we changed the initial $f(S,U_1,U_2)$ into constant value.

   \begin{eqnarray}
   \left\{
   \begin{array}{lll}
   \frac{du_{21}(t)}{dt} = -d_1u_{21}(t)+h_1u_{11}(t)-r_1u_{21}(t) \\
   ~\\
   \frac{du_{22}(t)}{dt} = -d_1u_{22}(t)+h_2u_{12}(t)-r_2u_{22}(t) \\
   ~\\
   \frac{du_{23}(t)}{dt} = -d_1u_{23}(t)+h_3u_{13}(t)-r_3u_{23}(t) \\
   ~\\
   \frac{du_{11}(t)}{dt} = -d_1u_{11}(t)+h_3u_{11}(t)-\alpha_1s_{1}(t) \\
   ~\\
   \frac{du_{12}(t)}{dt} = -d_1u_{12}(t)+h_3u_{12}(t)-\alpha_2s_{2}(t) \\
   ~\\
   \frac{du_{13}(t)}{dt} = -d_1u_{13}(t)+h_3u_{13}(t)-\alpha_3s_{3}(t).  \\
   \end{array}{}
   \right.
   \end{eqnarray}

The model that we developed can be written as

   \begin{eqnarray}
   \left\{
   \begin{array}{lll}
   \frac{ds_{1}(t)}{dt} =-\beta_1exp\{s_1(t)(-\sum_{k=1}^{3}(\alpha_{1k}s_k(t)+\alpha_{2k}u_{1k}(t)+\alpha_{3k}u_{2k}(t)))\}+c_1-ds_1(t)-\alpha_1s_1(t)\\
   ~\\
   \frac{ds_{2}(t)}{dt} = -\beta_2exp\{s_2(t)(-\sum_{k=1}^{3}(\alpha_{1k}s_k(t)+\alpha_{2k}u_{1k}(t)+\alpha_{3k}u_{2k}(t)))\}+c_2-ds_2(t)-\alpha_2s_2(t)\\
   ~\\
   \frac{ds_{3}(t)}{dt} = -\beta_3exp\{s_3(t)(-\sum_{k=1}^{3}(\alpha_{1k}s_k(t)+\alpha_{2k}u_{1k}(t)+\alpha_{3k}u_{2k}(t)))\}+c_3-ds_3(t)-\alpha_3s_3(t).\\
   \end{array}{}
   \right.
\end{eqnarray}

\begin{figure}[H]
      \centering
      \includegraphics[width=.8\textwidth]{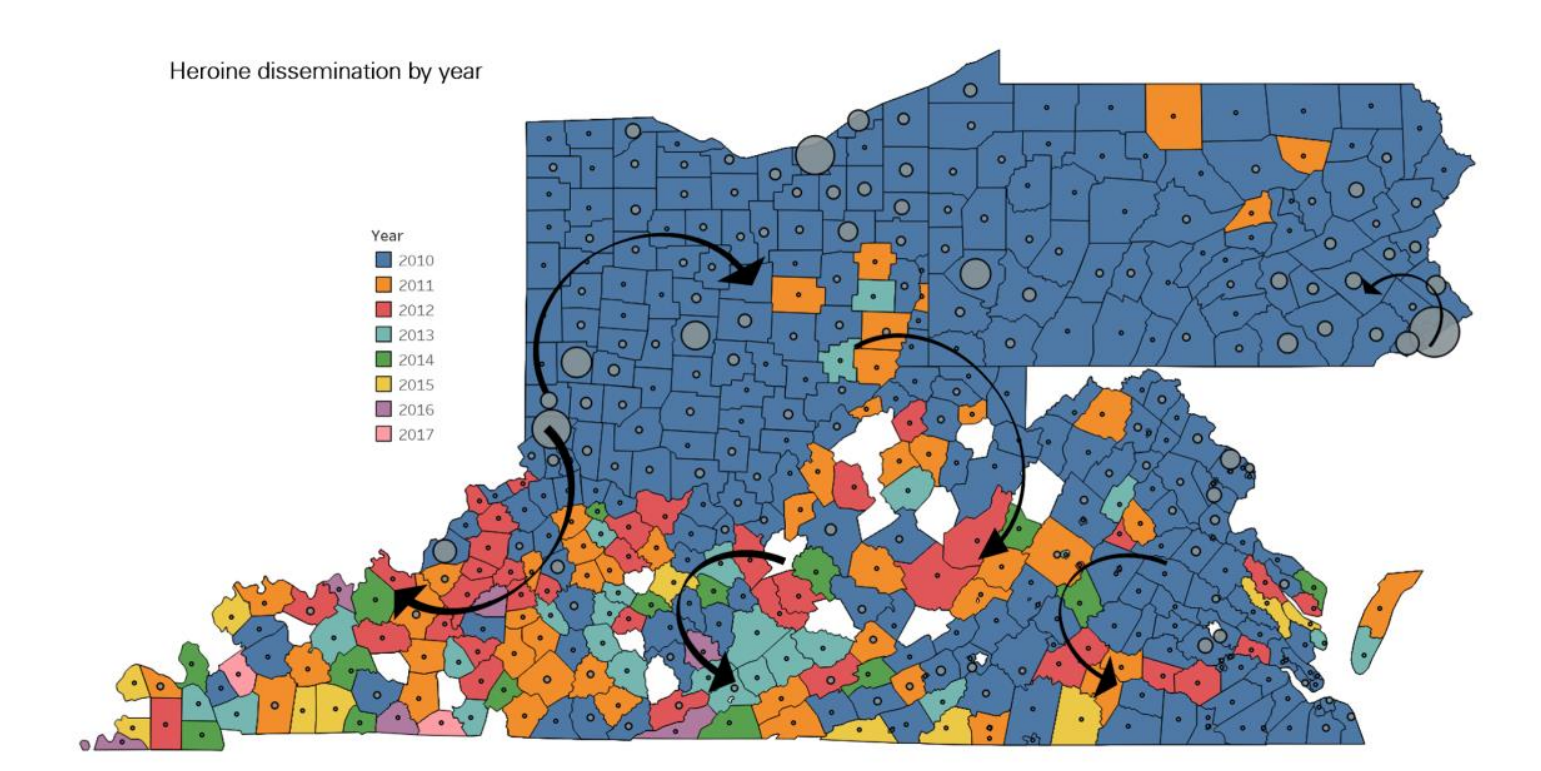}
      \caption{Opioid transmission among state and county}
\end{figure}

Despite the statistical gatherings, we may perform simulation on an autonomous internet system (AS) network. By analyzing the frequency of network centrality, the following places are shortlisted and identified as the source of opioid transmission.

\begin{table}[H]
\center
\begin{tabular}{@{}ccccccc@{}}
\toprule
State    & PA          & OH      & OH      & KY       & WV     & VA     \\ \midrule
County   & Philadelphia & Cuyahoga & Hamilton & Jefferson & Kanawha & Bedford \\
\midrule
Latitude & 40           & 41      & 39      & 38       & 38     & 37     \\
Longitude & -75         & -81     & -84     & -85      & -81    & -79    \\ \bottomrule
\end{tabular}
\end{table}

\section{Sensitivity analysis}
\subsection{Initial parameter values}

The initial parameter values we set in are:

$(c_1,c_2,c_3)=(1000,1000,1000)$

$(\beta_1,\beta_2,\beta_3)=(20,20,20)$

$(\alpha_1,\alpha_2,\alpha_3)=(0.5,0.5,0.5)$

$(\alpha_{11},\alpha_{12},\alpha_{13},\alpha_{21},\alpha_{22},\alpha_{23},\alpha_{31},\alpha_{32},\alpha_{33})=\protect\\
     (0,0,0,0,1/300000,1/300000,0,1/300000,1/300000)$

$(d,d_1)=(0.00001,0.01)$

$(h_1,h_2,h_3)=(0.2,0.2,0.2)$

$(r_1,r_2,r_3)=(0.2,0.2,0.2)$

\subsection{Testing results}

\begin{figure}[H]
      \centering
      \includegraphics[width=.8\textwidth]{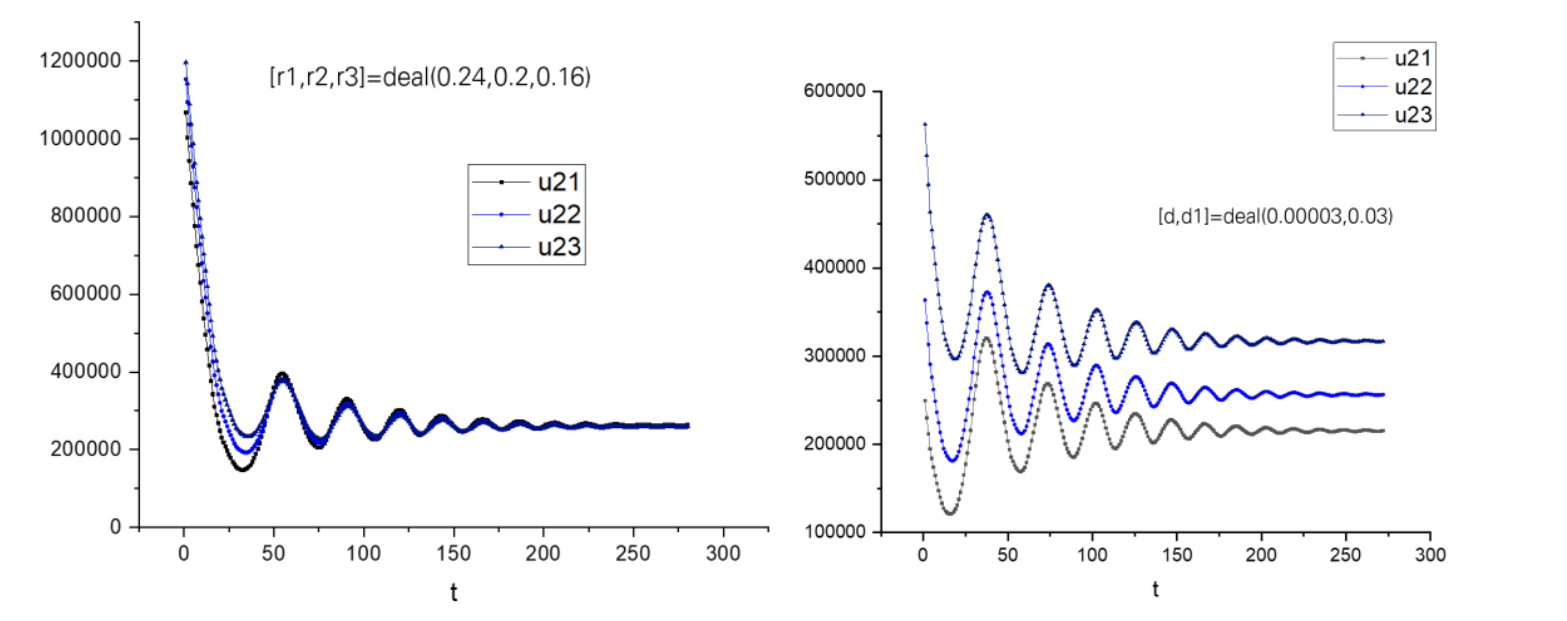}
      \caption{change parameters in $(r_1,r_2,r_3)=(0.24,0.20,0.16)$,$(d,d_1)=(3e-5,3e-2)$}
\end{figure}

\begin{figure}[H]
      \centering
      \includegraphics[width=.8\textwidth]{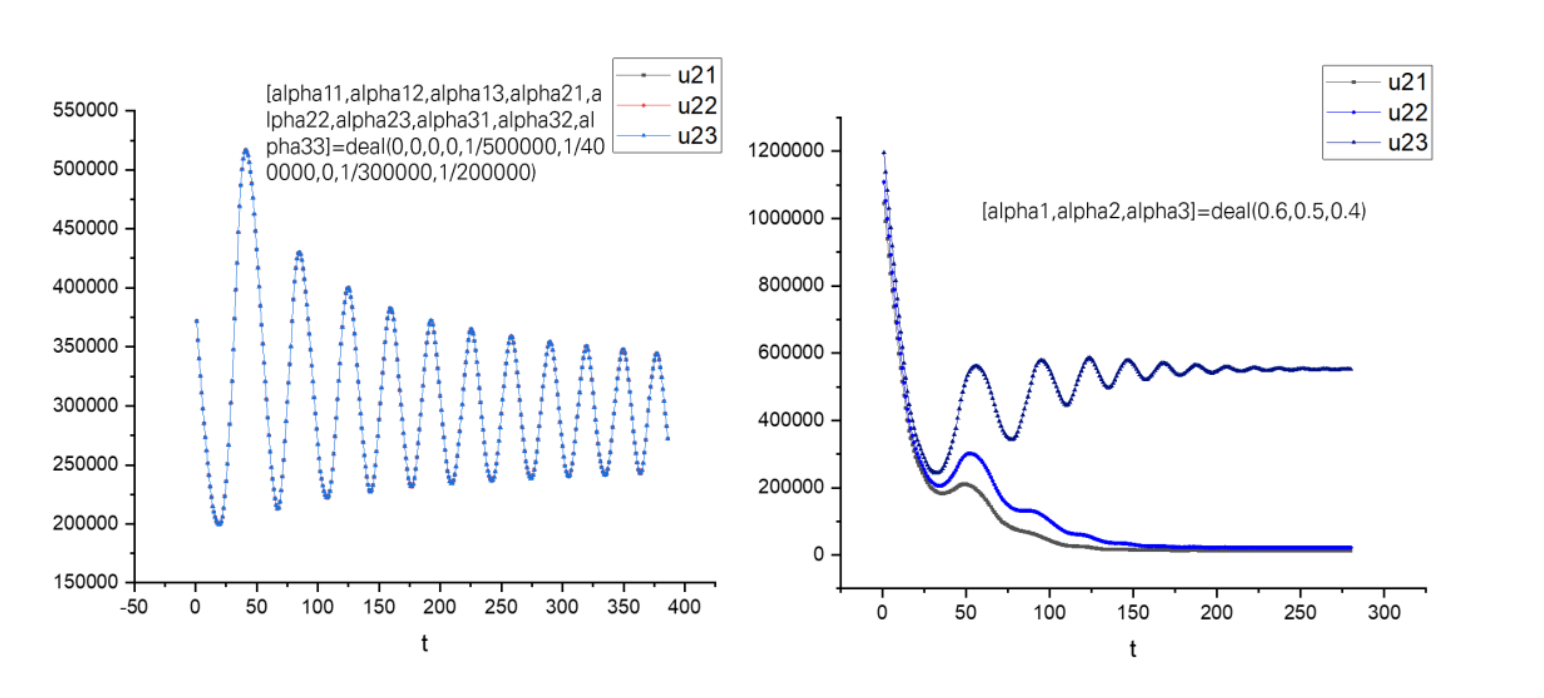}
      \caption{change parameters in $(\alpha_{11},\alpha_{12},\alpha_{13},\alpha_{21},\alpha_{22},\alpha_{23},\alpha_{31},\alpha_{32},\alpha_{33})=\protect\\
      (0,0,0,0,1/500000,1/400000,0,1/300000,1/200000)$,$(\alpha_1,\alpha_2,\alpha_3)=(0.6,0.5,0.4)$}
\end{figure}

\begin{figure}[H]
      \centering
      \includegraphics[width=.8\textwidth]{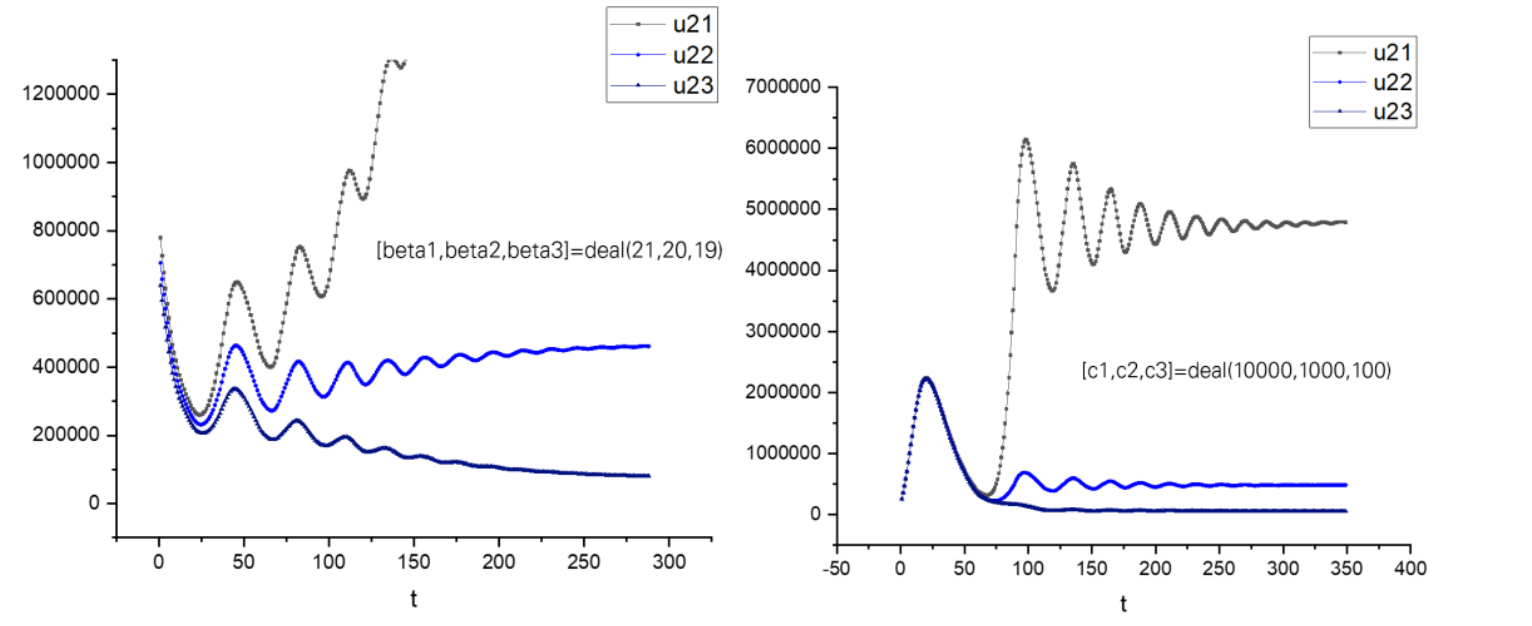}
      \caption{change parameters in $(\beta_1,\beta_2,\beta_3)=(21,20,19)$,$(c_1,c_2,c_3)=(1e4,1e3,1e2)$}
\end{figure}

\subsection{Implications}

\textbf{Adjustment of $(h_1,h_2,h_3)$}

Explanation: If the county increases the inspection of the infected person to a certain extent, although the number of infected people exposed in the county increases in the short term, it will still be similar to the number of infected people exposed in other counties in the long time. And the increase in the intensity of one county inspection is conducive to reducing the number of drug users in all counties.

\textbf{Adjustment of $(r_1,r_2,r_3)$}

Explanation: the same proportion of r adjustment within a specific range (for example, 4\% increase and a 4\% decrease), the effect of reduction is more considerable. This adjustment does not affect the trend of changes in the number of people. Therefore, the increase in the detoxification ability of a drug rehabilitation center can reduce the amount of addicts as a whole. Still, it does not change the fluctuation of the number of people and the time to reach the equilibrium point.

\textbf{Adjustment of $(d,d_1)$}

Explanation: If a drug makes the effect of addiction unchanged, but the mortality rate increases, it will reduce the number of addicts as a whole and reach the equilibrium point faster.

\textbf{Adjustment of $(\alpha_{ij})$}

Explanation: It shows that factor adjustment will not only change the trend of the number of addicts but also change the time to reach equilibrium.

\textbf{Adjustment of $(\alpha_i)$}

Explanation: An increase in the conversion rate of an urban addict will reduce the number of other urban addicts. And the same proportion of changes in the conversion rate (for example, the same difference of 0.1\%), the impact of the increase is higher. From the selected cases, it can be seen that the cities with the same conversion rate and a lower conversion rate are close to 0 when they reach the equilibrium point.

\textbf{Adjustment of $(\beta_i)$}

Explain: $\beta_i$ Is the main factor causing the impact. The change of this factor will cause considerable fluctuations in the number of addicted people. After the number of addicts reaches a certain level, it will increase exponentially. Therefore, the government should increase the intensity of news propaganda and reduce the impact of addicts with not addicted people.

\textbf{Adjustment of $(c_i)$}

Explanation: It shows that residents' happiness and education are also the main factors that affect them. The number of newly addicted people at each stage may not be high in absolute terms, but it will have a massive impact on the number of addicts at the end. Compared with the influence of $\beta_i$, the increase of $c_i$ will not break the balance, and ultimately the addicts There is still a balance point in quantity, but the time to reach the equilibrium point is delayed compared to normal.

\section{strengths \& weaknesses}
\subsection{strength}
\begin{itemize}
\item To predict the spread of drug addiction, we built the SUC model based on the SEIR model, and this model has a more detailed analysis of the behavior of several types of people, so it also has a great advantage in the degree of fit with the data.
\item     After processing the data, we classified the socio-economic data by the PCA method. We found 22 main influencing factors, which divided the county into three categories, which made the model more refined and then obtained various kinds of sensitivity analysis. The conclusion of the parameter adjustment can quantify the problem when making suggestions to the government.
\end{itemize}
\subsection{weakness}
\begin{itemize}
\item     There are many model parameters after refinement, although the results are more consistent, we need more data to calculate parameters, and qualitative analysis is more difficult.
\end{itemize}

\section{Conclusion}
\begin{itemize}
\item     The number of unexposed people who are not exposed is likely to exceed the number of knew addicted people, and the main influencing factor in changing drug addiction is that not addicted people are affected by the addicts. The government needs to increase the control of addicts and anti-drug news campaigns.
\item     The Relationship section is the most relevant part of data in drug transmission. We can make a reasonable assumption that high-quality relationships can reduce the possibility of drug addiction.
\end{itemize}

\end{document}